\documentclass[12pt,letterpaper]{article}
\usepackage[T1]{fontenc}
\usepackage[utf8]{inputenc}
\usepackage{lmodern,microtype}
\usepackage[margin=1in]{geometry}
\usepackage{amsmath,amssymb,amsthm,mathtools}
\usepackage{enumitem,setspace}
\usepackage[authoryear,round]{natbib}
\bibpunct{(}{)}{;}{a}{}{,}
\usepackage[hidelinks]{hyperref}
\newtheorem{theorem}{Theorem}
\newtheorem{proposition}{Proposition}
\newtheorem{lemma}{Lemma}
\newtheorem{corollary}{Corollary}
\theoremstyle{definition}
\newtheorem{assumption}{Assumption}

\newcommand{\dd}{\mathrm d}

\numberwithin{equation}{section}
\setlist{nosep,leftmargin=*}
\hypersetup{pdftitle={Reputation, Disclosure, and the Scope of Entry},pdfauthor={Rubik Khachatryan and Georgy Lukyanov}}
\title{Reputation, Disclosure, and the Scope of Entry\thanks{We thank Emiliano Catonini, Alla Friedman, Markus Gebauer, Vitalijs Jascisens, Ella Khromova, Steven Kivinen, Ekaterina Mitskevich, Konstantinos Shamruk, and Alexey Verenikin for helpful comments and conversations. All remaining errors are our own.}}
\author{Rubik Khachatryan\thanks{International College of Economics and Finance, HSE University, Moscow, Russia. Email: \href{mailto:rakhachatryan@edu.hse.ru}{rakhachatryan@edu.hse.ru}.}\and Georgy Lukyanov\thanks{Toulouse School of Economics, University of Toulouse Capitole, Toulouse, France. Corresponding author. Email: \href{mailto:georgy.lukyanov@tse-fr.eu}{georgy.lukyanov@tse-fr.eu}.}}
\date{}
\begin{document}
\maketitle
\begin{abstract}
We study how shared response resources affect reputation, entry, and disclosure. A later rival chooses whether to enter zero, one, or two markets. The incumbent then makes profitable cost-reducing investments. Shared costs make broader entry weaken the response in each market and can eliminate the incentive to enter only one. The incumbent's response cost is private information. Publishing its conduct in an earlier market encourages imitation by the less capable type, making a published response less informative and inducing broader entry. Our main result compares shared and separable response costs while holding all early and single-market payoffs fixed. We prove analytically that, on a nonempty open set of primitives with uniform setup costs, full publication maximizes total surplus under shared costs when publication is inexpensive, whereas no publication is optimal under separable costs. A capability audit is dominated in both economies. The reversal arises because shared costs strengthen the incentive for productive reputational investment. It does not require a disclosure-induced change in the available entry scales. Expected entry scope is unchanged across policies within each technology, so entry counts alone cannot evaluate disclosure.
\end{abstract}
\medskip
\section{Introduction}
\label{sec:introduction}

Publishing a firm's competitive history helps potential rivals assess the response they would face after entry. It also changes the firm's incentive to build that history. These effects depend on how the firm organizes its response across markets. When several markets draw on the same implementation resources, an entrant's decision to broaden its launch weakens the incumbent's response in markets already entered. How does this connection affect the value of making past conduct public?

We study a two-period Cournot model in which an incumbent privately knows its persistent cost of organizing a response. In an early, temporary market, it can undertake a standardized investment that lowers its production cost. A later rival observes an accurate record of this action with a given probability and chooses whether to enter zero, one, or two other markets. After entry, the incumbent invests in the contested markets and the firms compete in quantities. This final response is profitable without any further reputational benefit. The later entrant privately knows a sunk setup cost, which affects entry but not its operating technology.

The first result concerns entry scope. When response investments have a shared convex cost, entering an additional market reduces investment in the first. The entrant gains both a new market and a weaker response in its initial market. Under a stated payoff condition, this additional return removes the range of setup costs for which a one-market launch is optimal. We compare this outcome with a separable technology that preserves every early and single-market payoff. Without the interaction between responses, the one-market entry interval remains available at every belief. Individual firms may change scope discontinuously, but aggregate scope probabilities are continuous under atomless setup costs.

The second result concerns the meaning of a published response. The capable incumbent benefits from the early investment immediately; the less capable type makes a current private loss and responds only for its reputational return. Once publication is sufficiently frequent, this return induces imitation. More frequent publication then increases imitation and makes a response less informative about capability. The later rival consequently enters more markets conditional on that record. This is a comparison across publication regimes: within a regime, an observed response still indicates greater capability than an observed absence of response.

The welfare analysis distinguishes information from the conduct that information induces. A privately costly early response can raise total surplus by expanding output. Publishing conduct can induce this productive investment, whereas revealing capability before conduct removes the reputational incentive. Later entry also changes, and firms do not internalize its effects on consumers and the incumbent. We derive a decomposition of these two welfare effects and conditions for selecting a publication rule. With uniform setup costs, the comparison is analytical: welfare is quadratic in beliefs within an entry region, and its derivative along the reputation equilibrium has a constant direction of change within each region.

The main comparison holds every early and single-market payoff fixed and changes only the interaction between later responses. We prove that, on a nonempty open set of primitives with uniform setup costs, full publication is optimal under shared costs when publication is inexpensive, whereas no publication is optimal under separable costs. A capability audit is dominated in both economies. The proof constructs a family in which publication induces substantial productive imitation under shared costs but very little under separable costs. In the latter economy, the investment benefit is smaller than the later allocation loss. The reversal does not require one-market entry to disappear as disclosure increases. We also characterize the separable benchmark throughout the maintained domain: a capability audit lowers total surplus, and optimal conduct disclosure is either full publication or none, with an explicit threshold for the choice.

Under uniform setup costs, expected entry scope is invariant across the publication and audit policies within each technology. The invariance does not equate entry across technologies, and it does not make welfare invariant. Investment, setup costs, and the allocation of entry across incumbent types still matter. Thus mean entry alone cannot evaluate a disclosure rule. We also show analytically that arbitrarily small departures from the uniform distribution can give either sign for the unconditional entry effect.

The model is relevant to competition assessments in which operating adjustments use resources shared across markets and past conduct is incompletely observed. It identifies what such an application would have to establish: a current-profit reason to respond, a persistent private organizational cost, and records that are not already recoverable from other public information. It does not identify a particular industry's technology or evaluate a specific disclosure mandate.

\subsection{Related literature}

The reputational link between competitive encounters originates in \citet{krepswilson1982} and \citet{milgromroberts1982}; \citet{fudenbergkreps1987} study reputation against multiple opponents. \citet{srinivasan1991} analyzes cost signaling across markets, including entry undertaken for its signaling value. Here a later rival chooses its own scope, anticipating a profitable response by the incumbent. Shared response costs connect this choice to the incumbent's earlier reputation incentive. Other links between markets operate through uniform pricing \citep{fenglishuai2023} or commitment to a fighter-brand launch \citep{jost2023}. Our final response requires no prior commitment, distinguishing it from pre-entry capacity investment \citep{spence1977,dixit1980} and complete-information dynamic predation \citep{reyspiegelstahl2025}.

\citet{hwangkirby2000} provide the closest disclosure-and-entry comparison. In their Cournot model, more reliable cost disclosures change entry conditional on signals while leaving its ex ante probability constant under uniform-cost and interiority restrictions. Their welfare comparisons can favor either full or no disclosure. Our mean-scope identity has the same implication for the limits of entry counts as a welfare measure. The additional mechanism is productive reputational imitation followed by endogenous multimarket entry. Matching early and single-market payoffs isolates the interaction between later responses. In Hwang and Kirby, entry also reveals the entrant's private production cost. Our private setup cost is sunk before competition, so entry scope conveys no payoff-relevant information about operating technology.\footnote{Hwang and Kirby evaluate domestic welfare, excluding the foreign entrant's profit. We include all firms' profits and consumer surplus. Their disclosure parameter changes signal precision about exogenous costs; ours changes the frequency of an accurate record of a chosen action.}

Related work connects disclosure to endogenous investment. \citet{jiangxiong2025} study disclosure of costs determined by investment, and \citet{aryaetal2025} examine disclosure with product-quality investment and investment spillovers. \citet{ishinagioh2024} combine investment observability with congestion costs across markets. Both firms in their model operate in both markets; here the later rival chooses whether and where to enter, using beliefs formed from an earlier response. \citet{atanasov2024} studies disclosure incentives arising from information spillovers across a firm's segments. Our analysis links response technology, the value of reputation, and endogenous entry scope under common early and single-market payoffs.

Finally, observability changes incentives to acquire or demonstrate capability. The distinction between information and behavior is central to \citet{prat2005}, \citet{barisaac2012}, and \citet{barisaacdeb2021}; \citet{li2025} studies its role in policymaking and information provision. \citet{hoangruckes2017} connect disclosure, risk management, and entry. Here the continuation incentive comes from the later entrant's scope choice against an incumbent of persistent type.

Section \ref{sec:model} presents the model. Sections \ref{sec:response} and \ref{sec:reputation} solve the response, entry, and reputation decisions. Section \ref{sec:policy} compares disclosure policies, and Section \ref{sec:discussion} discusses competition implications and limits. The appendices give the policy derivations, the finite-market extension, and aggregation with nonuniform setup costs.

\section{The model}
\label{sec:model}

There are three firms: an incumbent $I$, an early rival $P$, and a later entrant $E$. The early rival operates in a temporary market; the later entrant chooses its own launch in two other markets. The incumbent's type $\theta\in\{H,L\}$ determines the cost of organizing a competitive response and remains the same across these encounters. Type $H$ has the lower cost. The incumbent knows its type, and the other firms assign prior probability $\mu\in(0,1)$ to $H$. Demand, base production costs, and physical capacity are public. The later entrant learns from a public record of the early action when that record is available. We first describe the later markets, whose entry decisions determine the incumbent's incentive to build reputation.

\subsection{The later markets}

The later entrant can serve two markets, indexed by $i=1,2$. Inverse demand in market $i$ is
\begin{equation}
 P_i=c_0+A_i-q_{Ii}-q_{Ei},\qquad A_1>A_2>0,
 \label{eq:demand}
\end{equation}
and its marginal production cost is $c_0>0$. The entrant privately draws a per-market setup cost $z$ from an atomless distribution $F$ on $[0,Z]$, independently of the incumbent's type and all observations. It chooses an entered set $S\subseteq\{1,2\}$ and incurs real cost $|S|z$.

After observing $S$, the incumbent chooses investment $x_i\in[0,c_0]$ on each entered market. Investment lowers its marginal cost from $c_0$ to $c_0-x_i$. These investments are observable before the firms choose quantities simultaneously. Their real cost is
\begin{equation}
 C_\theta(x;S)=\frac{\chi}{2}\sum_{i\in S}x_i^2+
 \frac{\gamma_\theta}{2}\left(\sum_{i\in S}x_i\right)^2,
 \qquad \chi>\frac89,\quad 0<\gamma_H<\gamma_L.
 \label{eq:cost}
\end{equation}
The first term is the cost of implementing investment locally. The second is the cost of drawing on resources used in both markets: an increase in $x_1$ raises the marginal cost of $x_2$, and conversely. We write $X_\theta(S)=\sum_{i\in S}x_{\theta i}(S)$ for total investment.

The shared cost describes the organization of simultaneous adjustments, rather than uncertainty about installed capacity. A lower $\gamma_\theta$ makes these adjustments less costly.\footnote{The quadratic specification yields an explicit solution. The relevant interaction is $\partial^2 C_\theta/\partial x_i\partial x_j=\gamma_\theta>0$ for $i\ne j$. This increasing marginal cost is distinct from a binding physical-capacity constraint.}

Investment is an adaptation to competition and is available only on entered markets. An unentered market retains marginal cost $c_0$ and monopoly output $A_i/2$. This restriction rules out a prior monopoly investment decision. If the same cost reduction were available before entry, that additional decision would have to be included in the game.

The entrant's setup cost is sunk before the incumbent invests. It changes the willingness to enter but has no effect on demand or marginal production cost. Learning about $z$ from the chosen scope therefore does not change the incumbent's operating response. This assumption separates the analysis from signaling by an entrant with private information about productivity or demand.

\subsection{The early market and public records}

The early rival has already entered a separate, one-period market with demand intercept $A_P$. The incumbent chooses between a standardized response $R$, reducing its marginal cost by $y$, and no response $A$. We assume
\begin{equation}
 0<y<\min\{A_P,c_0\}.
 \label{eq:earlymenu}
\end{equation}
The response costs $(\chi+\gamma_\theta)y^2/2$, as implied by the one-market version of \eqref{eq:cost}. As in the later markets, the rival's marginal cost is $c_0$; it observes the investment before Cournot competition. This observation within the early market is distinct from the public record available to the later entrant. The early rival's activity ends with this market episode. Its entry is given, and the binary response menu is a maintained restriction.\footnote{A standardized response may correspond to implementing a specified operating package rather than choosing an arbitrary level of investment. The subsequent allocation across entered markets is continuous. We do not claim that the binary early choice solves an unrestricted continuous signaling game.}

With probability $\tau\in[0,1]$, an independent verification process publishes an accurate record of the early action. Otherwise the record is empty. Publication is independent of the incumbent's type, its action, and the entrant's later cost draw. The later entrant observes this record and no other information about the early episode. An increase in $\tau$ therefore makes an accurate record more frequent; it does not improve the precision of a noisy signal.

A record may document whether a specified operating adjustment was implemented in a previous competitive episode. An empty record conveys no information because the same publication lottery applies to responses and absences of response. This interpretation requires that the later entrant cannot already infer the action from public prices, quantities, or other operating details. Freely observed and revealing prices would require a different information structure.

Resources become available again after the early episode. Only the organizational cost $\gamma_\theta$ persists. Early conduct thus affects later entry through beliefs, without using up resources needed in the later markets. The incumbent discounts later profit by $\delta\in(0,1]$, and we use the same weight for welfare. All firms maximize expected profit. We study perfect Bayesian equilibrium: investment and quantity choices are sequentially optimal, entry is optimal after each record, and beliefs satisfy Bayes' rule at records with positive probability.

The timing is as follows. The publication rule is fixed before the incumbent learns its type. The incumbent then chooses its early response, and the firms in that market compete in quantities. Verification produces a public record. After observing the record and its own setup cost, the later entrant chooses its markets. The incumbent allocates investment across the entered markets, and quantity competition follows. In Section \ref{sec:policy}, we compare publication rules and allow an independent public capability audit before the early action.

\section{Competitive responses and entry scope}
\label{sec:response}

\subsection{The incumbent's response}

We solve the game backwards. Suppose the later entrant has chosen a set $S$. For $x_i<A_i$, both firms produce positive quantities and Cournot competition gives
\begin{equation}
 q_{Ii}=\frac{A_i+2x_i}{3},\qquad q_{Ei}=\frac{A_i-x_i}{3},\qquad
 \Pi_{Ii}=\frac{(A_i+2x_i)^2}{9},\quad
 \Pi_{Ei}=\frac{(A_i-x_i)^2}{9}.
 \label{eq:cournot}
\end{equation}
Investment raises the incumbent's output and lowers its rival's output. It also raises total output. Let $b_i=4A_i/9$ and $d=\chi-8/9>0$. After omitting the constant zero-investment profit, the incumbent's objective on the interior region is
\begin{equation}
 \sum_{i\in S}b_i x_i-\frac d2\sum_{i\in S}x_i^2
 -\frac{\gamma_\theta}{2}\left(\sum_{i\in S}x_i\right)^2.
 \label{eq:responseobjective}
\end{equation}

\begin{lemma}
\label{prop:response}
For every entered set, the incumbent has a unique globally optimal response. If the following candidate satisfies $0<x_{\theta i}(S)<\min\{A_i,c_0\}$ for all $i\in S$, it is that response:
\begin{equation}
 X_\theta(S)=\frac{\sum_{i\in S}b_i}{d+|S|\gamma_\theta},\qquad
 x_{\theta i}(S)=\frac{b_i-\gamma_\theta X_\theta(S)}d.
 \label{eq:response}
\end{equation}
When the relevant candidates are interior, adding the other market strictly lowers investment in the initial market and strictly raises total investment. If both types' candidates are interior, the type with lower $\gamma_\theta$ invests more in each entered market.
\end{lemma}

\begin{proof}
The incumbent's operating profit has second derivative $8/9$ for $x_i<A_i$. If investment excludes the entrant, operating profit is $(A_i+x_i)^2/4$, with second derivative $1/2$. At $x_i=A_i$, profit is continuous and its first derivative falls from $4A_i/3$ to $A_i$. Subtracting local cost with $\chi>8/9$ makes each market's net return strictly concave over the whole feasible interval. Subtracting the convex shared cost preserves strict concavity in the vector. Continuity on the compact investment set gives existence; strict concavity gives uniqueness. An interior stationary candidate is therefore globally optimal, including against exclusionary deviations.

The first-order conditions are $b_i-dx_i-\gamma_\theta X=0$. Summing them gives \eqref{eq:response}. For $\gamma=\gamma_\theta$ and $i\ne j$,
\begin{align}
 x_{\theta i}(\{i\})-x_{\theta i}(\{i,j\})
 &=\frac{\gamma[(d+\gamma)b_j-\gamma b_i]}{d(d+\gamma)(d+2\gamma)},\label{eq:dilution}\\
 X_\theta(\{i,j\})-X_\theta(\{i\})
 &=\frac{(d+\gamma)b_j-\gamma b_i}{(d+\gamma)(d+2\gamma)}.\nonumber
\end{align}
The numerator is positive precisely because the pair response in market $j$ is positive. Finally,
\begin{equation}
 \frac{\partial x_{\theta i}(S)}{\partial\gamma_\theta}
 =-\frac{\sum_{j\in S}b_j}{(d+|S|\gamma_\theta)^2}<0.
 \label{eq:typeorder}
\end{equation}
\end{proof}

The response has a current-profit motive. At zero investment, its marginal return is $b_i>0$, so the incumbent invests even in the final period. An entrant anticipates this profitable investment when deciding where to launch. Shared implementation costs determine how the response is spread across markets.

Define the entrant's total operating profit $B_\theta(S)$ and the incumbent's loss from entry $\ell_\theta(S)$ relative to the unentered monopoly markets. Under interiority,
\begin{align}
 B_\theta(S)&=\sum_{i\in S}\frac{(A_i-x_{\theta i}(S))^2}{9},\label{eq:entrantprofit}\\
 \ell_\theta(S)&=\sum_{i\in S}\frac{5A_i^2}{36}
 -\frac12\sum_{i\in S}b_i x_{\theta i}(S).
 \label{eq:incumbentloss}
\end{align}
To obtain the second expression, monopoly profit is $A_i^2/4$, zero-investment duopoly profit is $A_i^2/9$, and the optimized value of \eqref{eq:responseobjective} is $\sum_i b_i x_i/2$. Thus a change in market demand affects entrant profit, incumbent loss, and the response incentive jointly. Losses need not be positive for arbitrary primitives; the conditions used below require their positivity and ordering explicitly.

\subsection{The entrant's scope decision}
\label{sec:entry}

Throughout this section, maintain the interiority condition in Lemma \ref{prop:response} for both types and every nonempty entered set. Suppose the later entrant's posterior probability of high capability is $p$. Its first question is which market to choose if it enters only one. From \eqref{eq:response},
\begin{equation}
 B_\theta(\{i\})=\frac{A_i^2}{9}
 \left(1-\frac{4}{9(d+\gamma_\theta)}\right)^2.
 \label{eq:singleton}
\end{equation}
Market 1 is therefore the preferred singleton for both types and at every posterior. We can write $S_0=\varnothing$, $S_1=\{1\}$, and $S_2=\{1,2\}$, abbreviating the associated profits and losses by $B_{\theta k}$ and $\ell_{\theta k}$. Quantities for $k=0$ are zero.

The entrant chooses the largest maximizer of
\begin{equation}
 U_k(p,z)=V_k(p)-kz,\qquad
 V_k(p)=B_{Lk}-pT_k,\qquad T_k=B_{Lk}-B_{Hk}.
 \label{eq:entry}
\end{equation}
The tie rule matters only for a zero-probability set of setup costs. The term $T_k$ measures how much less the entrant earns against a high-capability incumbent when it chooses scale $k$.

Define the return to the first market, the incremental return to the second, and their difference:
\begin{equation}
 a_1(p)=V_1(p),\qquad a_2(p)=V_2(p)-V_1(p),\qquad
 G(p)=a_1(p)-a_2(p).
 \label{eq:gap}
\end{equation}
The entrant chooses one market only if its setup cost is low enough to make the first market worthwhile and high enough to make the second unattractive. The following lemma characterizes this comparison and the effect of beliefs on the chosen scope.

\begin{lemma}
\label{prop:choice}
When $G(p)>0$, the entrant chooses two markets for $z\leq a_2(p)$, one market for $a_2(p)<z\leq a_1(p)$, and no entry for $z>a_1(p)$. When $G(p)\leq0$, it chooses two markets for $z\leq V_2(p)/2$ and no entry otherwise. All intervals are intersected with the cost support.

If $0=T_0\leq T_1\leq T_2$, chosen scope weakly increases as $p$ falls, for every setup cost. For atomless $F$, each scope probability $P_k(p)$ and mean scope $K(p)=\sum_k kP_k(p)$ are continuous in $p$.
\end{lemma}

\begin{proof}
One-market entry beats no entry if $z\leq V_1$, and it beats two-market entry if $z\geq V_2-V_1$. The larger-scale tie rule makes its selected interval $(a_2,a_1]$. This interval is empty when $G\leq0$, in which case comparison of the two endpoints gives the cutoff $V_2/2$. At $G=0$ the sole three-way tie is resolved in favor of two markets.

For the monotonicity claim, take $p'<p$ and two scales $j<k$. Moving to $p'$ increases $U_k-U_j$ by $(p-p')(T_k-T_j)\geq0$. If $k$ was chosen at $p$, the smaller scale cannot become strictly preferable at $p'$. If equality remains, the tie rule also excludes a switch to $j$.

When $G>0$, probabilities are $P_2=F(a_2)$, $P_1=F(a_1)-F(a_2)$, and $P_0=1-F(a_1)$. When $G\leq0$, they are $P_2=F(V_2/2)$, $P_1=0$, and $P_0=1-P_2$. Extend $F$ by zero and one outside its support. At $G=0$, both cutoffs equal $V_2/2$. Continuity of $F$ and the intercepts then proves the claim, including at the change of regime.
\end{proof}

The type ordering in Lemma \ref{prop:response} implies $T_1,T_2>0$. It does not imply $T_2\geq T_1$, which is a separate condition for monotone scope. The fixed best singleton follows from \eqref{eq:singleton}; with more general market heterogeneity, the identity of the preferred portfolio can itself change with beliefs.

Write $\Gamma_\theta=2B_{\theta1}-B_{\theta2}$. If
\begin{equation}
 \Gamma_L<0<\Gamma_H,
 \label{eq:crossing}
\end{equation}
then the one-market gap crosses zero at
\begin{equation}
 p^*=\frac{-\Gamma_L}{\Gamma_H-\Gamma_L}\in(0,1).
 \label{eq:pstar}
\end{equation}
Above $p^*$, there is a nonempty one-market cost interval, provided the cutoffs lie in the support. Below it, one-market entry is never chosen. The gap shrinks because $G'(p)=T_2-2T_1=\Gamma_H-\Gamma_L>0$: a decline in perceived capability raises the value of two-market entry by more than twice the increase in the singleton value.

The economic reason is the return that a broader launch creates in the first market. A low-capability incumbent finds it particularly costly to respond in both markets. Entering the second market can therefore raise the entrant's profit in the first by enough to outweigh the smaller size of the second. Against a high-capability incumbent, this benefit may be weaker. Beliefs then determine whether a narrow launch remains attractive.

\subsection{A matched separable benchmark}

To isolate the role of the shared resource, replace \eqref{eq:cost} by
\begin{equation}
 C^{\mathrm{sep}}_\theta(x;S)=\frac{\chi+\gamma_\theta}{2}\sum_{i\in S}x_i^2.
 \label{eq:separable}
\end{equation}
This changes only the interaction between investments in different markets. Every singleton response and every early-market payoff remain unchanged. With two-market entry, each investment is now chosen as if the other market were absent. Entrant operating profits are additive, and
\begin{equation}
 2B_\theta(\{1\})-B^{\mathrm{sep}}_\theta(\{1,2\})
 =B_\theta(\{1\})-B_\theta(\{2\})>0.
 \label{eq:separablegap}
\end{equation}
The one-market interval is therefore nonempty at every posterior when the cutoffs are interior. Reputation can still shift both thresholds. What disappears is the additional return to broad entry from reducing the response in the first market. We will use exactly this comparison in the welfare analysis, recomputing the reputation equilibrium under each technology.

For finitely many markets, the same comparison must also account for changes in the identity of the best portfolio at a given scale. Appendix \ref{app:finite} gives the exact choice intervals and shows why atomless costs preserve continuity of aggregate scope probabilities even when intermediate scales disappear.

\section{Reputation and disclosure}
\label{sec:reputation}

The later entrant would like to know which incumbent type it faces. The incumbent would like to be regarded as capable, since that reputation discourages entry. We now ask when the low type is willing to pay for an early response to obtain this reputational benefit, and how its incentive changes with publication.

\begin{assumption}
\label{ass:continuation}
For every nonempty entered set and both types, the candidate in \eqref{eq:response} is interior. The derived payoffs satisfy $0<\ell_{\theta1}<\ell_{\theta2}$ and $0<T_1<T_2$. The cost distribution has a continuous, strictly positive density on $[0,Z]$. For every $p\in[0,1]$, the pairwise cutoffs $V_1(p)$, $V_2(p)-V_1(p)$, and $V_2(p)/2$ lie in $(0,Z)$.
\end{assumption}

The assumptions give reputation a continuation value. Entry reduces incumbent profit, a broader launch causes a larger loss, and higher perceived capability deters entry. The support restriction ensures a positive mass of firms changes its entry decision when beliefs change. These conditions restrict the product-market primitives; they are not additional free continuation payoffs. In particular, demand and response costs jointly determine both firms' profits.

\begin{lemma}
\label{lem:loss}
Under Assumption \ref{ass:continuation}, the expected continuation loss
\begin{equation}
 \Lambda_\theta(p)=\sum_{k=0}^2\ell_{\theta k}P_k(p)
 \label{eq:loss}
\end{equation}
is continuous and strictly decreasing on $[0,1]$.
\end{lemma}

\begin{proof}
Continuity follows from Lemma \ref{prop:choice}. A lower $p$ weakly raises entry scope at every cost, and losses are strictly ordered by scope. The largest cost at which any entry occurs is $\max\{V_1(p),V_2(p)/2\}$. Both terms are strictly decreasing in $p$, so this cutoff strictly rises when $p$ falls. The interval between the old and new cutoffs has positive probability and changes from no entry to positive entry. Expected losses therefore strictly rise.
\end{proof}

The incumbent's current gain from choosing the standardized response in the early market is
\begin{equation}
 g_\theta=\frac{4A_Py}{9}-\frac{d+\gamma_\theta}{2}y^2.
 \label{eq:earlygain}
\end{equation}
Suppose $g_H>0>g_L$, and write $\kappa=-g_L>0$. The high type benefits from the response immediately. The low type would prefer not to respond in the absence of a reputational gain. In either case, the early rival continues to produce a positive quantity: after a response it supplies $(A_P-y)/3$ and earns $(A_P-y)^2/9$.

Consider strategies in which $H$ responds and $L$ responds with probability $\rho$. The overall response probability is $w=\mu+(1-\mu)\rho$. Bayes' rule gives
\begin{equation}
 p_R=\frac{\mu}{w},\qquad p_A=0,\qquad p_\varnothing=\mu.
 \label{eq:bayes}
\end{equation}
The no-response record reveals the low type when it occurs. An empty record does not change beliefs because verification is independent of the action and type. At $\tau=0$ the two published records are impossible; at $\tau=1$ the empty record is impossible. Beliefs at such records do not affect the outcome.

Define $D(p)=\Lambda_L(0)-\Lambda_L(p)$. This is the low type's reduction in expected entry loss when its reputation is $p$ instead of zero. Its total gain from responding is
\begin{equation}
 -\kappa+\tau\delta D(p_R).
 \label{eq:incentive}
\end{equation}
The low type pays $\kappa$ to make the early response. If the action is published, a response prevents the later entrant from identifying it as the low type, saving expected entry losses $\delta D(p_R)$. If no record appears, the entrant keeps the prior after either action. Only the published-record branch therefore matters for the incentive comparison.

\begin{proposition}
\label{prop:equilibrium}
Maintain Assumption \ref{ass:continuation}, $g_H>0>g_L$, and
\begin{equation}
 \delta D(\mu)<\kappa<\delta D(1).
 \label{eq:equilibriumconditions}
\end{equation}
Set $\tau_0=\kappa/[\delta D(1)]$ and $\tau_1=\kappa/[\delta D(\mu)]$, so that $0<\tau_0<1<\tau_1$. There is an equilibrium with $H$ responding surely and
\begin{equation}
 \rho(\tau)=0\quad\text{for }\tau\leq\tau_0,
 \qquad \tau\delta D(p_R(\tau))=\kappa\quad\text{for }\tau>\tau_0.
 \label{eq:mixing}
\end{equation}
Above $\tau_0$, the low type mixes with a unique $\rho\in(0,1)$. Its response probability is continuous and strictly increasing in $\tau$, and the response posterior is continuous and strictly decreasing. If, in addition,
\begin{equation}
 g_H>\delta[\Lambda_H(0)-\Lambda_H(1)],
 \label{eq:dominance}
\end{equation}
this equilibrium outcome is unique, apart from zero-probability entry ties and beliefs at impossible records.
\end{proposition}

\begin{proof}
As $\rho$ rises from zero to one, $p_R$ falls continuously from 1 to $\mu$. By Lemma \ref{lem:loss}, the gain $D(p_R)$ strictly falls. At $\rho=0$, \eqref{eq:incentive} is nonpositive precisely when $\tau\leq\tau_0$. At $\rho=1$, it is strictly negative even for $\tau=1$, by the left inequality in \eqref{eq:equilibriumconditions}. Hence there is a unique interior indifference point when $\tau>\tau_0$, and no positive mixing probability at $\tau=\tau_0$. The high type prefers response on this branch because its current gain is positive and the response record induces lower continuation losses than the no-response record. These strategies and Bayes' rule support the stated equilibrium.

The strict monotonicity and continuity of $D$ give a continuous inverse. An increase in $\tau$ requires a decrease in $D(p_R)$ to preserve indifference. Thus $p_R$ falls and $\rho$ rises.

For uniqueness, the greatest possible continuation disadvantage to $H$ from responding is at most $\delta[\Lambda_H(0)-\Lambda_H(1)]$. Condition \eqref{eq:dominance} therefore makes response strictly optimal for $H$ under any continuation beliefs. Pooling on response cannot be an equilibrium: the response posterior would be $\mu$, and even a deviation belief $p_A=0$ would provide the low type a reputation gain below $\kappa$. Pooling is thus excluded, and the no-response record reveals $L$ whenever verification has positive probability. The scalar argument then determines the low type's strategy. At $\tau=0$, the current payoff signs directly determine both actions.
\end{proof}

The upper bound on $\kappa$ allows sufficiently frequent publication to induce imitation. The lower bound prevents the low type from always responding, even at $\tau=1$. Hence publication first leaves the types separated and then induces mixing. The formal boundary $\tau_1>1$ records that pooling lies outside the feasible policy range.

Condition \eqref{eq:dominance} is sufficient for uniqueness. We maintain it in the welfare analysis so that policy comparisons do not depend on selecting among equilibria. Without this bound, the proposition establishes the equilibrium in which the high type always responds but need not exclude other outcomes.

\begin{corollary}
\label{cor:scope}
Under the conditions of Proposition \ref{prop:equilibrium}, raising verification within the mixing range weakly increases entry scope after a verified response for every fixed $z$, and strictly raises its expectation. Each conditional scope probability remains continuous. If \eqref{eq:crossing} holds and $p_R(1)<p^*<1$, one-market entry disappears at
\begin{equation}
 \tau^*=\frac{\kappa}{\delta D(p^*)}\in(\tau_0,1).
 \label{eq:taustar}
\end{equation}
\end{corollary}

\begin{proof}
The posterior falls strictly along the mixing range. Lemma \ref{prop:choice} then gives weakly greater scope at every cost, while the strictly moving participation cutoff in Lemma \ref{lem:loss} gives a positive-mass set of strict increases. Continuity follows from composition. The crossing is unique because $p_R$ is strictly decreasing, and substitution into the mixing equation gives \eqref{eq:taustar}.
\end{proof}

More frequent publication changes the composition of incumbents behind a response record: a larger fraction are low types. The response consequently becomes less effective at deterring a broad launch. This comparison is across publication regimes. Within any one regime, a response still indicates higher capability than an observed absence of response.

For a fixed setup cost that leads to entry under both policies, expected investment per entered market also falls in the mixing range. At a fixed scope, a lower probability of $H$ reduces expected investment. If scope rises from one market to two, investment in market 1 falls for each type; investment in market 2 is lower still because $A_2<A_1$. This is a comparison for the same entrant cost, rather than an average over a changing set of entering firms.

\subsection{Entry across all records}

The conditional expansion need not imply an increase in entry across all histories. In the uniform-cost case, when the pairwise cutoffs are interior,
\begin{equation}
 K(p)=\frac{V_2(p)}Z.
 \label{eq:affinemean}
\end{equation}
Indeed, if one-market entry is available, mean scope is $[G+2(V_2-V_1)]/Z=V_2/Z$. If it is unavailable, mean scope is $2(V_2/2)/Z$, giving the same expression. Since $V_2$ is affine in $p$ and $wp_R=\mu$,
\begin{equation}
 \overline K(\tau)=(1-\tau)K(\mu)+\tau[(1-w)K(0)+wK(p_R)]
 =K(\mu).
 \label{eq:meaninvariance}
\end{equation}
The expansion after a response is offset by changes in the frequencies and composition of records. This is an affine-payoff cancellation, not a general consequence of Bayes' rule.\footnote{\citet[Proposition 1]{hwangkirby2000} similarly obtain an invariant ex ante entry probability with uniform private production costs and interior entry cutoffs, despite changes in entry conditional on disclosure. Here the invariant object is the number of markets entered, and the distribution of records also responds to reputational imitation.} The distribution of scope and the allocation of entry can still change. Appendix~\ref{app:aggregation} proves that arbitrarily small smooth departures from the uniform distribution can give either sign for the unconditional effect.

\section{Disclosure and welfare}
\label{sec:policy}

We now compare two ways of making information available to the entrant. Publishing conduct preserves the incumbent's incentive to build reputation. Revealing capability before conduct removes that incentive, since the early response can no longer change what is known about the type. We evaluate these instruments by total surplus, taking account of both early investment and subsequent competition.

An information policy is chosen before the incumbent learns its type. With probability $\alpha\in[0,1]$, an independent audit publicly reveals the exact type before the early response. Audit selection is independent of type, and whether an audit occurs and its result are public before conduct. If no audit occurs, the prior remains $\mu$ and the early action is published with probability $\tau$. Expected real policy costs are $a\alpha+(1-\alpha)c\tau$, with $a,c\geq0$. We optimize over these two probabilities.\footnote{An audit requires verifiable information about organizational capability beyond the action record. Partial signals, type-dependent audit selection, an audit announced only after conduct, and selectively edited action records are outside this policy class.}

The objective includes consumer surplus and all firms' profits, net of real investment and setup costs. A welfare-maximizing disclosure authority is one interpretation. We do not model its institutional incentives or the incumbent's voluntary choice of what to disclose. A consumer-surplus objective or different distributional weights could select a different rule.

\subsection{Private and social response incentives}

In a contested market with an interior response, consumer surplus is $(2A_i+x_i)^2/18$. Adding operating profits gives
\begin{equation}
 CS_i+\Pi_{Ii}+\Pi_{Ei}=\frac{8A_i^2+8A_i x_i+11x_i^2}{18}.
 \label{eq:surplus}
\end{equation}
In the early market, moving from zero investment to $y$ increases consumer surplus by $2A_Py/9+y^2/18$ and changes the rival's profit by $-2A_Py/9+y^2/9$. Their sum is $y^2/6$. The total-surplus gain from response is therefore
\begin{equation}
 h_\theta=g_\theta+\frac{y^2}{6}.
 \label{eq:earlywelfare}
\end{equation}
The low type's privately costly response raises early total surplus whenever $\kappa<y^2/6$. The incumbent pays for investment but does not capture its full output benefit. Reputational imitation can therefore induce socially useful investment even though it lowers the incumbent's current profit.

For later entry, the surplus gain relative to the unentered monopoly markets, before setup costs, is
\begin{equation}
 \omega_\theta(S)=\sum_{i\in S}\left[\frac{5A_i^2}{72}
 +\frac{4A_i x_{\theta i}}9+
 \left(\frac{11}{18}-\frac\chi2\right)x_{\theta i}^2\right]
 -\frac{\gamma_\theta}{2}X_\theta^2.
 \label{eq:omega}
\end{equation}
Here monopoly surplus is $3A_i^2/8$, and response costs are already subtracted. At the optimal response, the first-order conditions give the useful equivalent expression
\begin{equation}
 \omega_\theta(S)=\sum_{i\in S}\left[\frac{5A_i^2}{72}
 +\frac{2A_i x_{\theta i}}9+\frac{x_{\theta i}^2}{6}\right].
 \label{eq:omegasimplified}
\end{equation}
It holds for both response technologies. For a fixed entered set, greater capability raises surplus as well as investment. Write
\begin{equation}
 W_\theta(p)=\int[\omega_\theta(S^*(p,z))-|S^*(p,z)|z]\,\dd F(z),
 \qquad W(p)=pW_H(p)+(1-p)W_L(p).
 \label{eq:W}
\end{equation}
The posterior determines the entry choice; the actual type determines the response and surplus. This distinction matters when the low type imitates the high type.

Up to constants common to the disclosure policies, gross welfare without an audit is
\begin{equation}
 \begin{split}
 \mathcal W_C(\tau)={}&\mu h_H+(1-\mu)\rho(\tau)h_L\\
 &+\delta\{(1-\tau)W(\mu)+\tau[(1-w)W(0)+wW(p_R)]\}.
 \end{split}
 \label{eq:wc}
\end{equation}
Under a full capability audit, each type acts on its current incentive: $H$ responds and $L$ does not. An action cannot change a type that is already known. Gross welfare is
\begin{equation}
 \mathcal W_F=\mu h_H+\delta[(1-\mu)W(0)+\mu W(1)].
 \label{eq:wf}
\end{equation}
Net expected welfare is
\begin{equation}
 \mathcal W(\alpha,\tau)=\alpha(\mathcal W_F-a)
 +(1-\alpha)[\mathcal W_C(\tau)-c\tau].
 \label{eq:regulator}
\end{equation}
The affine dependence on $\alpha$ follows from the timing: audit selection is independent and publicly resolved before conduct.

\subsection{Information and productive imitation}

Put
\begin{equation}
 A_F=\mathcal W_F-\mathcal W_C(0),\qquad
 \Delta=\mathcal W_C(1)-\mathcal W_C(0).
 \label{eq:policygains}
\end{equation}
The first quantity is the gross gain from revealing capability; the second is the gross gain from publishing conduct with certainty. Let $p_1=p_R(1)$ and $\rho_1=\rho(1)$. Equation \eqref{eq:wc} gives
\begin{equation}
 \Delta=(1-\mu)\rho_1h_L+
 \delta\left\{W(0)-W(\mu)+\frac{\mu}{p_1}[W(p_1)-W(0)]\right\}.
 \label{eq:gaindecomposition}
\end{equation}
The first term is the surplus from additional early investment. The second is the change in later surplus, including setup costs. The matched technologies have the same $h_L$ and $\kappa$. They generally have different $\rho_1$, $p_1$, and $W$, because the return to reputation and the consequences of entry depend on the later response technology. The decomposition is evaluated at each technology's own equilibrium.

In the separating range, publication changes information without changing early conduct:
\begin{equation}
 \mathcal W_C(\tau)=\mathcal W_C(0)+\tau A_F,
 \qquad 0\leq\tau\leq\tau_0.
 \label{eq:sepwelfare}
\end{equation}
Thus an adverse information effect can make low publication rates undesirable even when sufficiently frequent publication induces productive investment. Comparing full publication with no publication alone does not establish an optimum over all intermediate probabilities.

\begin{proposition}
\label{prop:policy}
Suppose the equilibrium in Proposition \ref{prop:equilibrium} is unique, $A_F<0$, and welfare is continuous and continuously differentiable on each of finitely many mixing intervals. If $\Delta>0$ and $\mathcal W_C'(\tau)>\Delta$ throughout those intervals, the optimal policy has no capability audit. Full publication is uniquely optimal for $0\leq c<\Delta$, and no publication is uniquely optimal for $c>\Delta$. At $c=\Delta$, only these two publication probabilities tie. These conclusions hold for every $a\geq0$.
\end{proposition}

\begin{proof}
The separating segment is strictly decreasing, also after costs. On the mixing range, integrate the strict derivative bound from any $\tau<1$ to 1 to obtain
\[
 \mathcal W_C(\tau)-\mathcal W_C(0)<\tau\Delta.
\]
If $c\geq\Delta$, every interior publication probability is therefore worse than no publication. If $c<\Delta$, net welfare is strictly increasing on each mixing interval, and full publication beats no publication by $\Delta-c>0$. Continuity joins the intervals. Finally, $\mathcal W_F-a<\mathcal W_C(0)$, so the affine expression \eqref{eq:regulator} excludes any positive audit probability at an optimum.
\end{proof}

These are sufficient conditions, rather than a claim that publication is desirable for all shared-cost technologies. They require the productive response to offset the initial information loss and bound the whole mixing path. Under uniform costs, Appendix \ref{app:policy} expresses the derivative in product-market payoffs. Its direction of change is constant in each entry region, so the bound can be established from the corresponding one-sided endpoint derivatives. When the bound does not hold, the same expressions identify any interior optimum analytically.

For a general technology, a capability audit need not be dominated. Equation \eqref{eq:regulator} implies that an optimum can always be chosen with $\alpha=0$ or $\alpha=1$: compare $\max_\tau\{\mathcal W_C(\tau)-c\tau\}$ with $\mathcal W_F-a$. A fractional audit probability is optimal only when these values tie. The strict condition $A_F<0$ is sufficient to rule out an audit, because no publication remains available without one.

\subsection{The matched separable benchmark}
\label{sec:separablepolicy}

With separable response costs, the entrant decides independently whether to serve each market. Both cutoffs are affine in its belief. Suppose setup costs are uniform and maintain the interiority, equilibrium, and uniqueness conditions for this technology. Define the single-market profits and type differences
\begin{equation}
 \beta_{\theta i}=B_\theta(\{i\}),\qquad
 t_i=\beta_{Li}-\beta_{Hi}>0,\qquad
 s_i=\omega_H(\{i\})-\omega_L(\{i\})>0.
 \label{eq:separablecoefficients}
\end{equation}
The last inequality follows from \eqref{eq:omegasimplified}. Put
\begin{equation}
 v=\frac1Z\sum_{i=1}^2\ell_L(\{i\})t_i>0,\qquad
 \eta=\frac1Z\sum_{i=1}^2t_i\left(s_i+\frac{t_i}{2}\right)>0.
 \label{eq:separablecoefficients2}
\end{equation}
Expected incumbent losses are affine in beliefs, with $D(p)=vp$. Welfare is a strictly concave quadratic, $W(p)=W(0)+\xi p-\eta p^2$, where the linear coefficient $\xi$ is immaterial for the comparisons below.

\begin{proposition}
\label{prop:separablepolicy}
Under the matched separable technology, uniform setup costs, and the maintained equilibrium and uniqueness conditions, let $\tau_0=\kappa/(\delta v)\in(\mu,1)$. A capability audit is strictly dominated by no disclosure for every $a\geq0$. Define
\begin{equation}
 \Delta_{\mathrm{sep}}=\mu\left[
 \frac{1-\tau_0}{\tau_0}h_L-\delta\eta(\tau_0-\mu)\right].
 \label{eq:separablegain}
\end{equation}
The optimal policy has $\alpha=0$ and full publication if $\Delta_{\mathrm{sep}}>c$, and no publication if $\Delta_{\mathrm{sep}}<c$. These optima are unique. If $\Delta_{\mathrm{sep}}=c\geq0$, only full and no publication tie. In particular, when publication is free, full publication is strictly preferred if and only if
\begin{equation}
 h_L>\frac{\delta\eta\tau_0(\tau_0-\mu)}{1-\tau_0}.
 \label{eq:separablethreshold}
\end{equation}
\end{proposition}

\begin{proof}
Let $\beta_i(p)=\beta_{Li}-pt_i$ and $\omega_i(p)=\omega_L(\{i\})+ps_i$. Additivity and direct integration give
\[
 W(p)=\frac1Z\sum_{i=1}^2
 \left[\omega_i(p)\beta_i(p)-\frac{\beta_i(p)^2}{2}\right].
\]
Its quadratic coefficient is $-\eta$. Hence
\begin{equation}
 A_F=\delta\{(1-\mu)W(0)+\mu W(1)-W(\mu)\}
 =-\delta\eta\mu(1-\mu)<0.
 \label{eq:separableaudit}
\end{equation}
Expected low-type losses are $\sum_i\ell_L(\{i\})\beta_i(p)/Z$, giving $D(p)=vp$ and $\mu<\tau_0<1$. In the mixing range, indifference and Bayes' rule give $p_R=\tau_0/\tau$ and $(1-\mu)\rho=\mu(\tau/\tau_0-1)$. Substituting the quadratic welfare function into \eqref{eq:wc} yields
\begin{equation}
 \mathcal W_C(\tau)-\mathcal W_C(0)
 =\mu h_L\left(\frac{\tau}{\tau_0}-1\right)
 -\delta\mu\eta(\tau_0-\mu\tau),\qquad \tau\geq\tau_0.
 \label{eq:separablepath}
\end{equation}
This segment is affine, as is the separating segment in \eqref{eq:sepwelfare}. Net welfare at $\tau_0$ is strictly below its value at zero. Its maximum must therefore occur at 0 or 1, and \eqref{eq:separablepath} at 1 gives \eqref{eq:separablegain}. The strict loss at $\tau_0$ also rules out an interior tie. Equation \eqref{eq:separableaudit} excludes any positive audit probability, and rearranging $\Delta_{\mathrm{sep}}>0$ gives \eqref{eq:separablethreshold}.
\end{proof}

The threshold separates two economically different effects. More capable incumbents generate greater surplus on a given contested market, but they also discourage entry. Revealing capability therefore directs entry away from the markets where a stronger response would create more surplus. Under the uniform distribution and separable technology, this allocation effect makes the audit costly in welfare terms. Conduct publication can compensate by inducing additional early investment. A positive $h_L$ is necessary for that compensation, but it is not sufficient.

The audit nevertheless weakly benefits the later entrant. Its expected net profit at belief $p$ is
\begin{equation}
 U(p)=\int\max_{k\in\{0,1,2\}}\{V_k(p)-kz\}\,\dd F(z).
 \label{eq:entrantvalue}
\end{equation}
This function is convex, being an integral of maxima of affine functions. Full capability information therefore gives at least as much expected entrant profit as any Bayesian signal about the same type. The private value of information does not include its effects on consumers, the incumbent, or early reputational investment.

\subsection{When shared resources reverse the publication rule}

The preceding conditions allow opposite policy choices under technologies with identical early and single-market payoffs. The next result establishes that such a comparison is feasible analytically.

\begin{theorem}
\label{thm:disclosure}
There is a nonempty open set of primitives within the uniform-cost family for which Assumption \ref{ass:continuation} and the equilibrium and uniqueness conditions of Proposition \ref{prop:equilibrium} hold in both matched technologies, and:
\begin{enumerate}[label=(\roman*)]
\item Under shared costs, no capability audit and full publication uniquely maximize total surplus when $0\leq c<\Delta_{\mathrm{sh}}$, where $\Delta_{\mathrm{sh}}>0$ is the gross full-publication gain in that economy. No audit and no publication are uniquely optimal for $c>\Delta_{\mathrm{sh}}$. Only these two publication choices tie at equality.
\item Under separable costs, no audit and no publication are uniquely optimal for every $c\geq0$.
\end{enumerate}
Both statements hold for every $a\geq0$. The set can be chosen so that one-market entry is never optimal under shared costs at any posterior. Thus a disclosure-induced disappearance of that scale is not necessary for the reversal.
\end{theorem}

\begin{proof}
Appendix \ref{app:construction} constructs a family of nearly symmetric markets and response-cost types indexed by $\varepsilon>0$. The early private loss tends to zero while the surplus gain from the low type's response remains positive. The setup-cost support is chosen so that the separable economy's imitation threshold tends to full publication. In the shared economy, the corresponding threshold tends to one half. The prior is below one half, so both equilibria remain in the maintained range.

Welfare is a strictly concave quadratic in beliefs and the low type's reputation benefit is linear in each economy. The mixing welfare paths are therefore affine. Under separable costs, imitation at full publication vanishes faster than the adverse allocation effect, giving $\Delta_{\mathrm{sep}}<0$. Under shared costs, the imitation gain has a positive limit and the later allocation effect vanishes, giving $\Delta_{\mathrm{sh}}>0$. The separating welfare slope is negative in both economies. Proposition \ref{prop:separablepolicy} applies to the separable economy; the shared affine path satisfies Proposition \ref{prop:policy}. All inequalities are strict for sufficiently small positive $\varepsilon$, and continuity gives an open neighborhood around each such economy. The appendix verifies the response, support, incentive, and uniqueness conditions directly.
\end{proof}

The reversal comes from different returns to reputation. Both types have the same early private and social payoffs across the matched technologies. Shared costs raise the value of avoiding a broad later launch sufficiently to induce more productive imitation. Under separable costs, the small investment gain does not recover the loss from directing entry toward less capable incumbents. The construction uses nearly symmetric markets and rare later entry. It establishes an open set with opposite policy choices, without asserting a uniform ranking or the prevalence of the reversal.

Finally, mean scope remains $K(\mu)$ under every publication and audit policy within either technology. On both audit and no-audit branches, the posterior has mean $\mu$ and \eqref{eq:affinemean} is affine. The welfare comparison consequently requires the investment and allocation terms, even when entry counts are unchanged.

\section{Competition implications and limits}
\label{sec:discussion}

The analysis gives three implications for assessing entry and disclosure. First, entry into an additional market can affect competition in an existing one even when demands and operating profits are otherwise independent. The link is the incumbent's allocation of response investment. An assessment based only on the added market would miss the weaker response in markets already contested. The prediction requires shared marginal implementation costs; common ownership by itself does not establish the mechanism.

Second, a response record has an endogenous meaning. Under the stated equilibrium conditions, making records more frequent encourages the less capable incumbent to respond, reduces capability conditional on a response, and raises entry scope after that record. These claims concern the same publication rule applied to responses and absences of response. Selective reporting would generally make an empty record informative and change the equilibrium. In an empirical application, differences in record availability would have to be distinguished from selection in which firms and actions are recorded. The model supplies conditional predictions, not an identification strategy.

Third, entry counts are insufficient for evaluating disclosure. In the uniform-cost case, mean entry is unchanged even though early investment and the allocation of later entry change. The relevant welfare comparison includes consumer surplus, investment costs, setup costs, and incumbent profit. A capability audit can help an entrant make privately better decisions while lowering total surplus. The separable benchmark establishes this distinction analytically; the shared-cost comparison requires the additional policy conditions stated above.

The response, scope, and reputation results allow general positive setup-cost densities under their maintained assumptions. Uniformity is used for the mean-scope identity, the explicit welfare formulas, and the open-set policy reversal in Theorem \ref{thm:disclosure}. Appendix \ref{app:aggregation} shows that small smooth departures from uniformity can give either sign for the unconditional entry effect. This does not by itself extend the optimal disclosure result to other distributions. With more markets, portfolio composition can also change at a fixed scale. Appendix \ref{app:finite} explains why scope probabilities can remain continuous while incumbent losses become discontinuous at such a composition switch.

Several assumptions delimit the application. The early market is temporary, resources become available again, and only organizational capability persists. A continuing early rival or durable investment would add a direct connection between periods. Early conduct is binary; later investment is continuous and available only after entry. Allowing monopoly investment would introduce another decision. The entrant's private setup cost is sunk before competition, so entry scope does not signal operating productivity. Finally, the action record is useful only if other public observations do not already reveal the action. These restrictions are needed to isolate learning from past conduct.

The disclosure comparison maximizes total surplus within a specified class of publication rules and capability audits. It does not characterize unrestricted information design, voluntary disclosure by the incumbent, or an authority with its own political or organizational incentives. Within that class, the main lesson is that response technology changes both the value of reputation and the conduct induced by publication. Evaluating information policy therefore requires more than its effect on the entrant's information or the average number of markets entered.

\clearpage
\appendix
\numberwithin{proposition}{section}
\numberwithin{lemma}{section}
\section{Analytical policy formulas}
\label{app:policy}

This appendix derives the welfare formulas used in Section \ref{sec:policy}. All coefficients are functions of demand, response costs, and the entry cutoffs. It also gives the candidate policies needed when the sufficient derivative condition in Proposition \ref{prop:policy} does not hold.

\subsection{A common representation of the two technologies}

For this calculation, introduce a parameter $\lambda\in[0,1]$ and write
\begin{equation}
 C_\theta^\lambda(x;S)=\frac{\chi+(1-\lambda)\gamma_\theta}{2}\sum_{i\in S}x_i^2+
 \frac{\lambda\gamma_\theta}{2}\left(\sum_{i\in S}x_i\right)^2.
 \label{eq:lambda}
\end{equation}
The shared and matched separable technologies correspond to $\lambda=1$ and $\lambda=0$. Every singleton cost equals $(\chi+\gamma_\theta)x^2/2$, independently of $\lambda$. Thus the interpolation preserves both singleton responses and the early gains. No monotonicity in $\lambda$ is asserted.

Let $d_\theta^\lambda=d+(1-\lambda)\gamma_\theta$ and $r_\theta^\lambda=\lambda\gamma_\theta$. On an interior entered set of size $k$, investment is
\begin{equation}
 X_\theta^\lambda(S)=\frac{\sum_{i\in S}b_i}{d_\theta^\lambda+kr_\theta^\lambda},
 \qquad x_{\theta i}^\lambda(S)=\frac{b_i-r_\theta^\lambda X_\theta^\lambda(S)}{d_\theta^\lambda}.
 \label{eq:lambdasolution}
\end{equation}
The formulas for $B_\theta$ and $\ell_\theta$ remain those in \eqref{eq:entrantprofit} and \eqref{eq:incumbentloss}. To compute $\omega_\theta$, replace $\chi$ in \eqref{eq:omega} by $\chi+(1-\lambda)\gamma_\theta$ and replace its final $\gamma_\theta$ by $\lambda\gamma_\theta$.

\subsection{Integration over setup costs}

Write $\omega_k(p)=p\omega_{Hk}+(1-p)\omega_{Lk}$ and maintain interior entry cutoffs with $F(z)=z/Z$. If $G(p)>0$, entry occurs in both markets for $z\leq a_2=V_2-V_1$ and in one market for $a_2<z\leq a_1=V_1$. Direct integration gives
\begin{align}
 W(p)&=\frac{\omega_1(p)G(p)+\omega_2(p)[V_2(p)-V_1(p)]-
 \tfrac12\{V_1(p)^2+[V_2(p)-V_1(p)]^2\}}{Z},\label{eq:Wthree}\\
 \Lambda_L(p)&=\frac{\ell_{L1}G(p)+\ell_{L2}[V_2(p)-V_1(p)]}{Z}.
 \label{eq:Lthree}
\end{align}
If $G(p)\leq0$, only zero and two-market entry are chosen, and
\begin{equation}
 W(p)=\frac{\omega_2(p)V_2(p)}{2Z}-\frac{V_2(p)^2}{4Z},
 \qquad \Lambda_L(p)=\frac{\ell_{L2}V_2(p)}{2Z}.
 \label{eq:Wtwo}
\end{equation}
The expressions agree at $G=0$. The welfare function is continuous and quadratic on each region, while the loss is continuous and affine. Global concavity of $W$ is neither assumed nor needed.

\subsection{Derivatives along the mixing path}

Within one entry region, let $W(p)=w_0+w_1p+w_2p^2$ and $D(p)=u+vp$, where $v>0$. Define
\begin{equation}
 A=\frac{\kappa}{\delta v},\quad B=\frac uv,\quad
 M=\mu h_L,\quad N=\delta\mu[w_0-W(0)].
 \label{eq:auxiliary}
\end{equation}
The mixing condition gives $p_R=A/\tau-B$, so $A-B\tau=\tau p_R>0$. Using $w=\mu/p_R$ in \eqref{eq:wc} yields
\begin{equation}
 \mathcal W_C(\tau)=C+L\tau+\frac{M\tau+N\tau^2}{A-B\tau},
 \label{eq:rational}
\end{equation}
where
\begin{align}
 C&=\mu(h_H-h_L)+\delta W(\mu)+\delta\mu w_2 A,\label{eq:Ccoefficient}\\
 L&=\delta[W(0)-W(\mu)+\mu w_1-\mu w_2 B].\label{eq:Lcoefficient}
\end{align}
Consequently,
\begin{align}
 \mathcal W_C'(\tau)&=L+
 \frac{MA+2NA\tau-NB\tau^2}{(A-B\tau)^2},\label{eq:derivative}\\
 \mathcal W_C''(\tau)&=\frac{2A(AN+BM)}{(A-B\tau)^3}.
 \label{eq:curvature}
\end{align}
The sign of the curvature is constant within the interval. Its smallest derivative therefore occurs at one of the endpoints. The one-sided derivatives on both sides of a region boundary must be checked.

If the same quadratic welfare formula holds over all posteriors and $D(p)=vp$, then $B=N=0$. Equation \eqref{eq:rational} is affine in $\tau$. Under the matched separable technology, this property holds throughout the maintained interior domain, because one-market entry is available at every posterior. Under shared costs it holds in any entry region whose polynomial extends to $p=0$: there $u=0$ and $w_0=W(0)$. If $G$ changes sign, the other region need not have this affine property.

\subsection{An analytical family with opposite publication choices}
\label{app:construction}

\begin{proof}[Proof of Theorem \ref{thm:disclosure}]
Fix $A,A_P,\bar\kappa>0$, $d>8/9$, $0<a_H<a_L$, $0<\mu<1/2$, $\delta\in(0,1)$, and $c_0>\max\{A,A_P\}$. For sufficiently small $\varepsilon>0$, set
\begin{equation}
 A_1=A,\quad A_2=A(1-\varepsilon^2),\quad
 \gamma_\theta=a_\theta\varepsilon,\quad
 \chi=d+\frac89,\quad \kappa=\bar\kappa\varepsilon^2.
 \label{eq:analyticfamily}
\end{equation}
Let $b_P=4A_P/9$ and choose the early response
\begin{equation}
 y=\frac{b_P+\sqrt{b_P^2+2(d+a_L\varepsilon)\bar\kappa\varepsilon^2}}
 {d+a_L\varepsilon}.
 \label{eq:analyticy}
\end{equation}
Then $g_L=-\kappa$ exactly and
\begin{equation}
 y\longrightarrow y_0=\frac{8A_P}{9d}<A_P,\qquad
 h_L\longrightarrow h_0=\frac{y_0^2}{6}>0,\qquad
 g_H=\frac{(a_L-a_H)y_0^2}{2}\varepsilon+O(\varepsilon^2)>0.
 \label{eq:analyticearly}
\end{equation}
Thus the early response remains interior and productive, while its low-type private loss is of smaller order than the high type's current gain.

For the later markets define
\begin{equation}
 \begin{split}
 f(s)&=\frac19\left(1-\frac4{9s}\right)^2,\qquad
 e=A^2\left(\frac5{36}-\frac8{81d}\right)>0,\\
 c_B&=A^2 f'(d)(a_L-a_H)>0,\qquad x_0=\frac{4A}{9d},\\
 c_\omega&=\left(\frac{2A}{9}+\frac{x_0}{3}\right)
 \frac{4A}{9d^2}(a_L-a_H)>0.
 \end{split}
 \label{eq:analyticcoefficients}
\end{equation}
Singleton operating profit is $A_i^2f(d+\gamma_\theta)$. Smoothness around $\varepsilon=0$ and \eqref{eq:omegasimplified} give, for both markets,
\begin{equation}
 \ell_L(\{i\})=e+O(\varepsilon),\qquad
 t_i=c_B\varepsilon+O(\varepsilon^2),\qquad
 s_i=c_\omega\varepsilon+O(\varepsilon^2).
 \label{eq:analyticexpansions}
\end{equation}
Write $\mathcal L_{\mathrm{sep}}=\sum_i\ell_L(\{i\})t_i$ and choose the common uniform setup-cost support as
\begin{equation}
 Z=\frac{\delta\mathcal L_{\mathrm{sep}}(1-\varepsilon^4)}{\bar\kappa\varepsilon^2}.
 \label{eq:analyticZ}
\end{equation}
Since $\mathcal L_{\mathrm{sep}}=2ec_B\varepsilon+O(\varepsilon^2)$, $Z$ is positive and tends to infinity at order $\varepsilon^{-1}$. By construction, the separable imitation threshold is exactly $\tau_{0,\mathrm{sep}}=1-\varepsilon^4$.

To examine shared costs, first take the symmetric-market limit. A response to two-market entry then uses the coefficient $d+2\gamma_\theta$ in place of $d+\gamma_\theta$. The market-size difference in \eqref{eq:analyticfamily} is only of order $\varepsilon^2$. Consequently,
\begin{equation}
 \Gamma_\theta=-2A^2 f'(d)a_\theta\varepsilon+O(\varepsilon^2)<0,
 \qquad T_{2,\mathrm{sh}}=4c_B\varepsilon+O(\varepsilon^2).
 \label{eq:analyticshared}
\end{equation}
Both endpoint gaps are negative, so $G(p)<0$ throughout $[0,1]$. Also $\ell_{L2,\mathrm{sh}}=2e+O(\varepsilon)$. Define $\mathcal L_{\mathrm{sh}}=\ell_{L2,\mathrm{sh}}T_{2,\mathrm{sh}}/2$. Then
\begin{equation}
 \frac{\mathcal L_{\mathrm{sh}}}{\mathcal L_{\mathrm{sep}}}\longrightarrow2,
 \qquad
 \tau_{0,\mathrm{sh}}=(1-\varepsilon^4)
 \frac{\mathcal L_{\mathrm{sep}}}{\mathcal L_{\mathrm{sh}}}
 \longrightarrow\frac12.
 \label{eq:analyticthresholds}
\end{equation}

All continuation conditions hold for sufficiently small positive $\varepsilon$. Each response tends to $x_0\in(0,A)$, below $c_0$; singleton losses tend to $e>0$ and pair losses to $2e$. The singleton type exposure is $c_B\varepsilon+O(\varepsilon^2)$, whereas pair exposure is twice as large to first order under separable costs and four times as large under shared costs. Thus $0<T_1<T_2$ in both economies. The three pairwise entry cutoffs have positive finite limits while $Z$ diverges, giving interior cutoffs for all $p$. The uniform density is continuous and strictly positive.

In both technologies $D(p)$ is linear through zero. Equations \eqref{eq:analyticZ} and \eqref{eq:analyticthresholds}, together with $\mu<1/2$, give $\mu<\tau_0<1$, which is equivalent to the two low-type incentive inequalities. The high type's maximum discounted reputation loss is of order $\varepsilon^2$: its type-dependent cutoff differences are of order $\varepsilon$, losses are bounded, and $1/Z$ is of order $\varepsilon$. Its current response gain is strictly positive of order $\varepsilon$ by \eqref{eq:analyticearly}. The uniqueness inequality therefore holds as well.

It remains to compare welfare. From \eqref{eq:separablecoefficients2} and \eqref{eq:analyticexpansions},
\begin{equation}
 \eta_{\mathrm{sep}}=\bar\eta\varepsilon^3+O(\varepsilon^4),\qquad
 \bar\eta=\frac{\bar\kappa(c_\omega+c_B/2)}{\delta e}>0.
 \label{eq:analyticeta}
\end{equation}
Using $\tau_{0,\mathrm{sep}}=1-\varepsilon^4$ in \eqref{eq:separablegain} gives
\begin{equation}
 \Delta_{\mathrm{sep}}=-\mu\delta\bar\eta(1-\mu)\varepsilon^3
 +O(\varepsilon^4)<0.
 \label{eq:analyticsepdelta}
\end{equation}
The productive imitation gain is only of order $\varepsilon^4$, so it cannot offset this allocation loss.

Under shared costs, \eqref{eq:Wtwo} holds globally. Let $s_{2,\mathrm{sh}}=\omega_{H2,\mathrm{sh}}-\omega_{L2,\mathrm{sh}}>0$. Its welfare function has negative quadratic coefficient
\begin{equation}
 -\eta_{\mathrm{sh}}=-\frac1Z\left[
 \frac{s_{2,\mathrm{sh}}T_{2,\mathrm{sh}}}{2}
 +\frac{T_{2,\mathrm{sh}}^2}{4}\right]<0,
 \qquad \eta_{\mathrm{sh}}=O(\varepsilon^3).
 \label{eq:analyticsharedeta}
\end{equation}
Since $D(p)$ is linear, the derivation of \eqref{eq:separablepath} applies with this technology's own $\eta$ and $\tau_0$. Hence
\begin{equation}
 \Delta_{\mathrm{sh}}=\mu\left[
 \frac{1-\tau_{0,\mathrm{sh}}}{\tau_{0,\mathrm{sh}}}h_L
 -\delta\eta_{\mathrm{sh}}(\tau_{0,\mathrm{sh}}-\mu)\right]
 \longrightarrow\mu h_0>0.
 \label{eq:analyticshareddelta}
\end{equation}
In either economy, $A_F=-\delta\eta\mu(1-\mu)<0$, and the mixing welfare segment is affine. Under shared costs its slope is $(\Delta_{\mathrm{sh}}-\tau_0 A_F)/(1-\tau_0)>\Delta_{\mathrm{sh}}$, establishing the condition in Proposition \ref{prop:policy}. Proposition \ref{prop:separablepolicy} gives no publication under separable costs.

For each sufficiently small positive $\varepsilon$, all response, loss, cutoff, incentive, uniqueness, gap, and welfare-sign inequalities used above are strict. They are continuous in the primitives. The separable quadratic and affine properties hold structurally, and the shared ones persist while both $\Gamma_\theta$ remain negative. The same ranking therefore holds in a nonempty open neighborhood of each constructed economy within the uniform-cost family.
\end{proof}

\subsection{Optimal policies outside the sufficient endpoint condition}

Maintain uniform setup costs, interior cutoffs, and the unique equilibrium. The separating segment is affine. Each mixing region has the form \eqref{eq:rational}, and subtracting the linear publication cost does not change its curvature. In a strictly convex region, an interior point cannot maximize net welfare. In a strictly concave region, there is at most one stationary point, determined by
\begin{equation}
 (L-c)(A-B\tau)^2+MA+2NA\tau-NB\tau^2=0.
 \label{eq:stationarypolicy}
\end{equation}
Only a root inside the relevant mixing region is a candidate. On an affine region, endpoints suffice; if the slope equals $c$, the entire interval ties internally.

Consequently, a global maximum can be obtained by comparing $\tau=0$, $\tau=1$, $\tau_0$, any interior publication rate at which $G(p_R)=0$, and the stationary points in strictly concave mixing regions. At a change of entry region, use the polynomial appropriate to each side. Continuity ensures that no other value is missed. If a zero-slope affine interval reaches the global maximum, all its points are also optimal. This describes the possible ties without asserting that an interior optimum or a policy reversal occurs for every response technology.

Because $G$ is affine, it has at most one interior zero unless it vanishes identically. In the latter case, the zero-or-two-market formula applies throughout. Finally, compare the best conduct-policy value with $\mathcal W_F-a$ using \eqref{eq:regulator}. Thus the stated policy class can be evaluated from analytical expressions, without discretizing the publication probability.

\section{Entry scope with finitely many markets}
\label{app:finite}

This appendix extends the scale results to finitely many markets and states the additional conditions needed for monotonicity and the reputation equilibrium.

Let the available markets be a finite set $\mathcal N$. For any portfolio $S\subseteq\mathcal N$, let $B_\theta(S)$ denote the entrant's operating profit under type $\theta$, after the incumbent's optimal response. Its best operating profit at scale $k$ is
\begin{equation}
 v_k(p)=\max_{|S|=k}\{pB_H(S)+(1-p)B_L(S)\},\qquad v_0(p)=0.
 \label{eq:finitevalues}
\end{equation}
Each $v_k$ is continuous and piecewise affine, but need not be globally affine because the best portfolio can change. Choose the largest maximizing scale in $v_k(p)-kz$. Define
\begin{equation}
 l_k(p)=\max_{j>k}\frac{v_j(p)-v_k(p)}{j-k},\qquad
 u_k(p)=\min_{j<k}\frac{v_k(p)-v_j(p)}{k-j},
 \label{eq:finitebounds}
\end{equation}
with the empty maximum equal to $-\infty$ and the empty minimum equal to $+\infty$. The selected-scale cell is
\begin{equation}
 I_k(p)=(l_k(p),u_k(p)]\cap[0,Z].
 \label{eq:finitecell}
\end{equation}
The half-open cell implements largest-scale selection: a tie with a larger scale excludes $k$, whereas a tie with a smaller scale includes it. A cell with $l_k\geq u_k$ is empty. Scales unsupported by the upper boundary of $(k,v_k)$ are never uniquely optimal; \eqref{eq:finitecell} also restricts costs to $[0,Z]$.

\begin{proposition}
\label{prop:finitecontinuity}
For an atomless distribution of setup costs, every scale probability induced by \eqref{eq:finitevalues} is continuous in $p$.
\end{proposition}

\begin{proof}
Fix $p$. Distinct scales have different payoff slopes in $z$, so they can tie only at finitely many costs. Outside those costs, the selected scale is the unique maximizing scale and remains so in a neighborhood of $p$, by continuity of all $v_k$. The indicators of each selected scale therefore converge almost surely as posteriors approach $p$. Dominated convergence gives continuity of their probabilities. Ties between different portfolios of the same scale do not affect this argument.
\end{proof}

A sufficient condition for entry scope to increase as $p$ falls is that for every $k>j$ and $p'<p$,
\begin{equation}
 v_k(p')-v_j(p')\geq v_k(p)-v_j(p).
 \label{eq:finitemonotone}
\end{equation}
To verify it, suppose scale $k$ is selected at $p$ but a smaller scale $j$ is selected at $p'$. The optimality inequalities and \eqref{eq:finitemonotone} imply equality throughout; largest-scale selection then contradicts the choice of $j$ at $p'$. When each scale uses a fixed portfolio and $v_k(p)=B_{Lk}-pT_k$, nondecreasing $T_k$ is sufficient. No such ordering follows from portfolio optimization alone.

For completeness, consider a finite reduction from $p_1$ to $p_2<p_1$ and two scales $0\leq h<h'\leq|\mathcal N|$. Their pairwise intersection is
\begin{equation}
 \zeta_{h'h}(p)=\frac{v_{h'}(p)-v_h(p)}{h'-h}.
 \label{eq:finiteintersection}
\end{equation}
Up to finitely many tie costs, the costs switching exactly from $h$ to $h'$ are
\begin{equation}
 \bigl(\zeta_{h'h}(p_1),\zeta_{h'h}(p_2)\bigr)
 \cap I_h(p_1)^{\circ}\cap I_{h'}(p_2)^{\circ}.
 \label{eq:finiteswitch}
\end{equation}
Both optimality cells are needed to exclude a third scale. Under \eqref{eq:finitemonotone}, positive-probability switches strictly increase mean scope, since no costs switch downward. Yet a scale disappearing at a single posterior cannot create an aggregate jump, by Proposition \ref{prop:finitecontinuity}.

Continuity of scale probabilities does not imply continuity of incumbent losses. If equally large portfolios exchange optimality, all costs choosing that scale can switch composition together; different losses across those portfolios can then cause a jump. The two-market model avoids this issue because market 1 is the strictly best singleton. More generally, the reputation equilibrium needs additional conditions on composition switches.

\section{Aggregation across published and unpublished records}
\label{app:aggregation}

Let $K(p)$ be expected entry scope under an atomless cost distribution satisfying the continuation assumptions. Along the equilibrium with $H$ responding surely, write $w=\mu+(1-\mu)\rho$, so $p_R=\mu/w$. The difference between mean scope with a verified record and mean scope without a record is
\begin{equation}
 J(w)=(1-w)K(0)+wK(\mu/w)-K(\mu).
 \label{eq:J}
\end{equation}
It can also be written as
\begin{equation}
 J(w)=(1-w)[K(0)-K(\mu)]-w[K(\mu)-K(p_R)].
 \label{eq:Jlevels}
\end{equation}
If $K$ is decreasing, both bracketed differences are nonnegative; their difference need not be. At a differentiability point,
\begin{equation}
 J'(w)=K(p_R)-K(0)-p_RK'(p_R).
 \label{eq:Jderivative}
\end{equation}
The level difference is nonpositive and the derivative term nonnegative: imitation changes both the frequency and content of response records.

Unconditional mean scope and its derivative are
\begin{align}
 \overline K(\tau)&=K(\mu)+\tau J(w(\tau)),\label{eq:generalmean}\\
 \overline K'(\tau)&=J(w(\tau))+\tau J'(w(\tau))w'(\tau).
 \label{eq:generalderivative}
\end{align}
The first term in \eqref{eq:generalderivative} holds the experiment fixed; the second captures its change through imitation. Neither has a general sign.

The absence of a general sign can be established analytically near any uniform-cost economy satisfying the strict maintained equilibrium conditions. Keep the demand and response primitives fixed and perturb only the setup-cost distribution:
\begin{equation}
 F_\varepsilon(z)=(1-\varepsilon)\frac zZ+
 \varepsilon\left(\frac zZ\right)^2,\qquad |\varepsilon|<1.
 \label{eq:nonuniformfamily}
\end{equation}
Its density is at least $(1-|\varepsilon|)/Z>0$. Write $G_+(p)=\max\{G(p),0\}$ and define
\begin{equation}
 Q(p)=\frac{V_2(p)^2+G_+(p)^2}{2Z^2}.
 \label{eq:aggregationQ}
\end{equation}
Using the entry intervals gives, in both regimes,
\begin{equation}
 K_\varepsilon(p)=(1-\varepsilon)\frac{V_2(p)}Z+\varepsilon Q(p).
 \label{eq:perturbedmean}
\end{equation}
When $G>0$, the quadratic term is $[a_1^2+a_2^2]/Z^2$; when $G\leq0$, it is $V_2^2/(2Z^2)$. This proves \eqref{eq:perturbedmean} across the boundary as well.

\begin{proposition}
\label{prop:nonuniformsign}
Consider any uniform-cost economy satisfying the strict conditions of Assumption \ref{ass:continuation} and Proposition \ref{prop:equilibrium}, including uniqueness. There is a neighborhood of $\varepsilon=0$ in \eqref{eq:nonuniformfamily} in which these conditions continue to hold. On a common separating interval with positive length,
\begin{equation}
 \overline K_\varepsilon'(\tau)=\varepsilon\left[
 (1-\mu)Q(0)+\mu Q(1)-Q(\mu)\right],
 \label{eq:nonuniformsigns}
\end{equation}
where the bracket is strictly positive. Thus arbitrarily small smooth departures from the uniform distribution yield either sign for the unconditional entry effect.
\end{proposition}

\begin{proof}
The operating outcomes and entry cutoffs do not depend on $F$. Expected losses depend continuously on $\varepsilon$, uniformly in $p$, since each is a finite sum of fixed losses times the continuous cutoff probabilities. The strict incentive and uniqueness inequalities therefore persist for sufficiently small $|\varepsilon|$. The separation threshold also remains strictly positive, so a common separating interval exists. On that interval $\rho=0$, $w=\mu$, and $p_R=1$. Substitution of \eqref{eq:perturbedmean} into \eqref{eq:generalmean} cancels the affine $V_2$ terms and gives \eqref{eq:nonuniformsigns}.

The function $V_2(p)^2$ is strictly convex because its quadratic coefficient is $T_2^2>0$. The squared positive part of the affine function $G$ is convex. Their sum makes $Q$ strictly convex, so the bracket is positive for $0<\mu<1$. Taking either sign of a sufficiently small nonzero $\varepsilon$ proves the result.
\end{proof}

This result concerns the unconditional entry derivative on the separating interval. It neither signs that derivative throughout the mixing range nor extends the uniform-cost optimal disclosure result.

At an entry-region boundary, $p_R$ falls as $\tau$ rises. A right derivative in $\tau$ therefore uses a left derivative in $p$. The mixing condition gives
\begin{equation}
 p'_{R,+}(\tau)=-\frac{D(p_R)}{\tau D'_{-}(p_R)},\qquad
 p'_{R,-}(\tau)=-\frac{D(p_R)}{\tau D'_{+}(p_R)},
 \label{eq:onesided}
\end{equation}
provided the one-sided derivatives of $D$ exist and are positive. Apply the same chain rule to \eqref{eq:generalderivative}. At the separation boundary, the derivative from below has $w'=0$. Levels remain continuous.

Under uniform costs and interior cutoffs, affine $K(p)=V_2(p)/Z$ implies $J(w)=0$. Equations \eqref{eq:generalmean}--\eqref{eq:generalderivative} then give the exact zero effect in \eqref{eq:meaninvariance}, including at kinks. Both the cost distribution and the entry geometry are needed.

\end{document}